\documentclass[a4paper,11pt]{article}
\usepackage{jheppub} 
\usepackage{lineno}
\usepackage{comment}
\usepackage{float}

\title{\boldmath Resummation-scale uncertainties in PDF determinations}

\author[a,b]{The xFitter Developers' team: Hamed Abdolmaleki,}
\author[c]{Valerio Bertone,}
\author[d,e]{Giuseppe Bozzi,}
\author[f,g]{Francesco Giuli,}
\author[h]{Alexander Glazov,}
\author[i]{Valentina Guglielmi,}
\author[j]{Francesco Hautmann,}
\author[k]{Fred Olness,}
\author[l]{Pavel Starovoitov,}
\author[m]{Oleksandr Zenaiev}

\affiliation[a]{School of Physics, Institute for Research in Fundamental Sciences (IPM), Tehran, Iran}
\affiliation[b]{Department of Physics, Malayer University, Malayer, Iran}
\affiliation[c]{IRFU, CEA, Universit\'{e} Paris-Saclay, F-91191 Gif-sur-Yvette, France}
\affiliation[d]{Dipartimento di Fisica, Universit\'{a} di Cagliari, Cittadella Universitaria, I-09042 Monserrato}
\affiliation[e]{INFN, Sezione di Cagliari, Cittadella Universitaria, I-09042 Monserrato}
\affiliation[f]{Universit\'{a} degli Studi Link, Via del Casale di S. Pio V, 44, 00165 Rome, Italy}
\affiliation[g]{INFN Sezione di Roma Tor Vergata, Via della Ricerca Scientifica, 1, 00133, Rome, Italy}
\affiliation[h]{Deutsches Elektronen-Synchrotron DESY, D-22607 Hamburg}
\affiliation[i]{Physik-Institut, Universit\"{a}t Z\"{u}rich, Z\"{u}rich, Switzerland}
\affiliation[j]{Theoretical Physics Department, University of Oxford, Oxford OX1 3PU}
\affiliation[k]{SMU Physics, Dallas, USA}
\affiliation[l]{University of Sharjah, 27272 Sharjah, UAE}
\affiliation[m]{II. Institut f\"{u}r Theoretische Physik, Universit\"{a}t Hamburg Luruper Chaussee 149, D-22761 Hamburg, Germany}

\emailAdd{Hamed.Abdolmaleki@desy.de} 
\emailAdd{valerio.bertone@cea.fr}
\emailAdd{giuseppe.bozzi@unica.it}
\emailAdd{francesco.giuli@cern.ch}
\emailAdd{alexander.glazov@desy.de}
\emailAdd{valentina.guglielmi@physik.uzh.ch}
\emailAdd{francesco.hautmann@physics.ox.ac.uk}
\emailAdd{olness@smu.edu}
\emailAdd{pstarovoitov@sharjah.ac.ae}
\emailAdd{oleksandr.zenaiev@desy.de}

\abstract{We investigate the impact of resummation-scale variations on parton distribution function (PDF) determinations. These variations introduce an additional source of theoretical uncertainty related to the evolution of PDFs and strong coupling $\alpha_s$ that has never previously been accounted for in PDF fits. Using the open-source \texttt{xFitter} framework, we quantify this uncertainty in fits to HERA deep-inelastic scattering data and examine its implications for predictions of top-quark pair production at the LHC. We show that the resulting resummation-scale uncertainties have a sizable impact on LHC observables, highlighting the importance of including this source of uncertainty in precision QCD analyses. 
}

\begin{document}
\maketitle
\flushbottom

\section{Introduction}

Precision phenomenology at present hadron colliders (LHC, RHIC)  
as well as at future experimental facilities 
(HL-LHC~\cite{Azzi:2019yne,LHeC:2020van}, 
EIC~\cite{Proceedings:2020eah}, FCC~\cite{FCC:2018byv}) 
requires accurate control of the theoretical uncertainties in the evolution of parton distribution functions (PDFs). While theoretical uncertainties associated with the 
standard variations of the factorisation and renormalisation scales in the QCD 
hard-scattering factorisation formulas are routinely assessed, 
additional sources of uncertainty arise from the perturbative truncation of renormalisation group 
equations (RGEs). RGE uncertainties are due to unknown higher orders in the 
perturbative expansion of the kernels of the evolution equations. While 
factorisation/renormalisation scale variations take into account 
theory uncertainties from QCD dynamics around the scale of the 
hard scattering, RGE uncertainties depend on the effects of QCD dynamics 
building up over the whole evolution span from the evolution starting scale to the hard-scattering scale.

Recently, it has been pointed out~\cite{Bertone:2022sso,Bertone:2024snr}
that such RGE uncertainties can be taken into
account by introducing the method of the so-called resummation scales and 
their variations, in a manner analogous to what is done in the QCD
literature on soft-gluon or Sudakov resummations. 
 This is based on extending the resummation-scale 
 technique from the case of 
 double-logarithmic resummations to the case of the single-logarithmic 
 resummation of collinear logarithms, embodied in the RGE fulfilled 
 by PDFs.  
The theoretical uncertainties that are taken into account  
by variations of the resummation scale  reflect ambiguities 
in truncating RGEs at a given order of nominal accuracy in 
perturbation theory, owing to formally subleading orders. That is, 
they correspond to differences between RGEs
which are entirely equivalent from the viewpoint of the nominal perturbative accuracy, yet may differ by  subleading terms. The method has also been developed for the case of the evolution of transverse momentum dependent parton distributions~\cite{Angeles-Martinez:2015sea}, and similar techniques are applied in resummed calculations 
of the Drell-Yan transverse momentum distribution~\cite{Camarda:2019zyx}.  

While  RGE effects  and  resummation-scale variations are 
routinely included in  QCD calculations of threshold or 
transverse momentum resummation (see, e.g.,~\cite{Catani:2003zt,Bozzi:2005wk}),  
they are usually not taken into account in the literature on 
PDF determinations. 
The purpose of this work 
is to perform for the first time a determination of PDFs from fits to collider data  
by including RGE theoretical uncertainties through the use of 
resummation-scale variations. We aim to assess the impact of 
resummation-scale effects on 
the physics of PDF extractions, and the role of the resulting 
PDF uncertainties in the case of observables at the LHC and at 
future colliders. 

We limit ourselves to carrying out 
PDF fits to deep-inelastic scattering (DIS) data from HERA~\cite{H1:2015ubc}. 
 The same methodology with resummation-scale variations can 
  be applied to global fits including hadron-hadron collider data. 
We perform PDF fits at next-to-next-to-leading order (NNLO) in 
perturbative QCD by implementing the resummation-scale 
formalism~\cite{Bertone:2024snr} as implemented in {\tt APFEL++}~\cite{Bertone:2013vaa, Bertone:2017gds}
in the open-source \texttt{xFitter} framework~\cite{xFitter:2022zjb,Alekhin:2014irh}.  
The resulting PDFs are then used to make predictions for top-quark 
production at the LHC and FCC in NNLO QCD via the \texttt{Hathor} package~\cite{Aliev:2010zk}.

 The paper is structured as follows. In Sec.~\ref{sec-resscale} we 
 briefly recap the resummation-scale method to take into account 
 RGE theoretical uncertainties. In Sec.~\ref{sec-setup} we describe 
 the basic set-up for the DIS fits. In Sec.~\ref{sec-pdf} we present 
 the results for PDFs including resummation-scale uncertainties. 
 In Sec.~\ref{sec-lhc} we study the impact on top-quark production 
 at the LHC and FCC.  In Sec.~\ref{sec-conc} we give conclusions.

\section{Resummation scale in RGE evolution}
\label{sec-resscale} 

We consider RGEs of the general form fulfilled, for instance, by PDFs and the strong coupling $\alpha_S$,  
 that is, 
\begin{equation}
\label{rge1} 
	\frac{d \ln R(\mu)}{d \ln \mu} = \Gamma(\alpha_S(\mu)),
\end{equation}
where $R$ is a renormalised quantity, $\mu$ is the renormalisation 
scale, and $\Gamma$ is 
the appropriate anomalous dimension, which admits a perturbative expansion in powers of $\alpha_S$, 
\begin{equation}
\label{perturb1}
	\Gamma(\alpha_S) = \sum_{n=0}^{\infty} \Gamma_n \alpha_S^{n+1}.
\end{equation}

At fixed perturbative order, i.e.,  when the series is truncated at some finite order, different prescriptions for solving the RGE in 
Eq.~(\ref{rge1}) exist which differ by subleading terms. From a perturbative viewpoint, such solutions are all equivalent and differ only by terms beyond the nominal accuracy, related to the limited knowledge of the anomalous dimension. Therefore, their spread can be used to estimate the truncation uncertainty.

When fitting PDFs to experimental data at different mass scales, RGEs are used to evolve PDFs to the data scales, where they are convoluted with appropriate partonic cross section to produce predictions. Therefore, in this context, theoretical predictions become affected by RGE uncertainties, which thus need to be accounted for to reliably estimate PDF uncertainties.

The goal of the resummation-scale technique is to provide a consistent procedure for evaluating these theoretical systematic uncertainties.
The detailed application of this technique takes different forms according to whether the RGE solution is obtained analytically or numerically. Both cases are fully treated in Ref.~\cite{Bertone:2024snr}.  
Here we focus the discussion on the 
case of the numerical solution of RGEs, which we use for the PDF extractions presented in this paper.
 
The resummation-scale technique applied to numerical RGE solutions is based on variations of Eq.~(\ref{rge1}), which all correspond to truncating the anomalous dimension
$\Gamma$ in Eq.~(\ref{perturb1}) at order $k$, but differ by (arbitrary) contributions of order higher than $k$.
For each variation, a numerical solution of the RGE is obtained.
The numerical differences between these solutions provide a 
method to evaluate the systematic uncertainties of RGE.
Since the anomalous dimension $\Gamma$
in Eq.~(\ref{rge1}) is a perturbative quantity, the customary way to
generate higher-order terms is to vary the scale at which the coupling is computed, and re-expand around the new scale truncating at order $k$. As an example, take the anomalous
dimension $\Gamma$ in Eq.~(\ref{perturb1}) truncated at order $k=1$,
\begin{equation}
\Gamma =\Gamma_0 \alpha_S(\mu) + \Gamma_1
\alpha_S^2(\mu).
\label{eq:NLOanomalousdimension}
\end{equation}
Next, we solve the RGE for the
running coupling $\alpha_S$ truncated to one-loop accuracy,
\begin{equation}
\label{rge-alphas}
\frac{d\ln \alpha_S(\mu)}{d\ln \mu} = \beta_0
\alpha_S(\mu),
\end{equation}
to evolve it from scale $\xi\mu$ to the scale $\mu$, where the factor $\xi$ is dubbed \textit{resummation-scale parameter}. The result can be written in the well-known form 
\begin{equation}
\label{as-ximu}
\alpha_S (\mu ) = \frac{\alpha_S (\xi\mu)}
{1 + \beta_0 \alpha_S (\xi\mu) \ln \xi}.
\end{equation}
Provided that $\xi$ is not too different from unity (a standard range used in practical applications is $1/2 \leq \xi \leq 2$), one can safely expand the solution above to $O(\alpha_S^2)$, obtaining
\begin{equation}
\alpha_S(\mu) = \alpha_S(\xi\mu) - \alpha_S^2 (\xi\mu)
\beta_0\ln\xi +\mathcal{O}(\alpha_S^3).
\label{eq:alphasNLOexp}
\end{equation}
Plugging this into Eq.~(\ref{eq:NLOanomalousdimension}) and retaining terms up to the nominal $O(\alpha_S^2)$ gives
\begin{equation}
\Gamma = \Gamma_0 \alpha_S(\xi\mu)+\alpha_S^2 (\xi\mu)
\left[\Gamma_1-\beta_0 \Gamma_0 \ln\xi\right]
+\mathcal{O}(\alpha_S^3).
\label{eq:NLOanomalousdimensionShifted}
\end{equation}
For $\xi\neq1$, Eqs.~(\ref{eq:NLOanomalousdimension})
and~(\ref{eq:NLOanomalousdimensionShifted}) are formally equivalent (same perturbative accuracy) but numerically different. Therefore, solving Eq.~(\ref{rge1}) with either of them produces perturbatively equivalent but numerically different solutions. As a consequence, variations of $\xi$ around $\xi = 1$ provide a method to estimate RGE uncertainties.

The technique just described 
can be applied to any finite order of truncation. In this paper, we apply it at order $k = 2$, i.e., NNLO, for the RGEs fulfilled 
by PDFs (and $\alpha_S$\footnote{We take into account the discontinuities of $\alpha_S$ across flavor thresholds, starting at NNLO.}). 
That is, we solve the DGLAP evolution 
equations~\cite{Gribov:1972ri,Altarelli:1977zs,Dokshitzer:1977sg}, using the
three-loop splitting 
functions~\cite{Moch:2004pa,Vogt:2004mw}, with different values of $\xi$ to assess the associated RGE systematic uncertainty. 
A detailed discussion of RGE uncertainties at different perturbative orders for various quantities, including the strong coupling and DIS structure functions, can be found in Ref.~\cite{Bertone:2024snr}.

We observe that the RGE uncertainties are cumulative, that is, they arise along the whole interval  spanned by  the evolution from the initial scale to the hard scattering scale.
This may be contrasted with the variations of the factorisation and renormalisation scales, which only involve variations about the hard scattering scale. 
This makes resummation-scale and factorisation/renormalisation-scale uncertainties quantitatively and qualitatively 
different. While the latter uncertainties are routinely taken into account in modern PDF fits~\cite{Ball:2022hsh,Ball:2021leu,NNPDF:2019ubu,Bailey:2020ooq,Hou:2019efy,Alekhin:2017kpj,NNPDF:2024dpb,NNPDF:2024nan,McGowan:2022nag,Cridge:2023ryv,Cridge:2023ozx}, the purpose of this work is to assess, for the first time, the impact of the former.




\section{PDF fit setup}
\label{sec-setup}

The analysis is performed using inclusive HERA DIS data~\cite{H1:2015ubc}, initially following the setup of Ref.~\cite{Bonvini:2019wxf}. The fit is carried out at NNLO in QCD using the FONLL-C scheme for heavy flavours~\cite{Forte:2010ta}, with $\alpha_S(M_Z) = 0.118$, the charm-quark mass $m_c = 1.46~\mathrm{GeV}$, and the beauty-quark mass $m_b = 4.5~\mathrm{GeV}$. The calculations use the {\tt APFEL++} library~\cite{Bertone:2013vaa, Bertone:2017gds} interfaced in {\tt xFitter}. The resummation scale is controlled by the $\xi$ parameter (see previous section) which is provided through the input file. It is possible to set different resummation scales for the evolution of PDFs and $\alpha_S$.

The PDFs parametrised at the starting scale of QCD evolution as a function of the partonic momentum fraction $x$ are the gluon distribution, $xg(x)$, the valence-quark distributions, $xu_v(x)$ and $xd_v(x)$, and the $u$-type and $d$-type antiquark distributions $x\bar{U}(x)$ and $x\bar{D}(x)$ (also referred to as sea ($\Sigma$) quark distributions):
\begin{align}
	xg(x)&=A_gx^{B_g}(1-x)^{C_g}(1+F_g\textrm{log}(x)+{G_g}\textrm{log}^2(x)), \\
	xu_v(x)&=A_{u_v}x^{B_{u_v}}(1-x)^{C_{u_v}}(1+E_{u_v}x^2+F_{u_v}\textrm{log}(x)+G_{u_v}\textrm{log}^2(x)), \\
	xd_v(x)&=A_{d_v}x^{B_{d_v}}(1-x)^{C_{d_v}}, \\
	x\bar{U}(x)&=A_{\bar{D}}x^{B_{\bar{D}}}(1-x)^{C_{\bar{U}}}(1+D_{\bar{U}}x+F_{\bar{D}}\textrm{log}(x)), \\
	x\bar{D}(x)&=A_{\bar{D}}x^{B_{\bar{D}}}(1-x)^{C_{\bar{D}}}(1+D_{\bar{D}}x+F_{\bar{D}}\textrm{log}(x)).
\end{align}
Here $A$, $B$, $C$, $D$, $E$, $F$, $G$ are free parameters of the fit, with some of them being constrained by the momentum and flavour sum rules. Furthermore, $B_{\bar{U}}$ and $B_{\bar{D}}$ are set equal, and $A_{\bar{U}}=A_{\bar{D}}(1-f_s)$ where $f_s=0.4$ is an $x$-independent strange suppression fraction~\cite{H1:2015ubc}. The strange quark and antiquark distribution is expressed as $x{\bar{s}(x)}=f_sx{\bar{D}(x)}$. This parametrisation has $17$ free parameters in total, and provides sufficiently flexible shapes and improved description of the HERA data, compared to the 14-parameter fit used in the HERAPDF2.0 analysis~\cite{H1:2015ubc}. We do not assign separate model and parametrisation uncertainties, since their impact is substantially smaller than the increased fit uncertainties obtained with the more flexible parametrisation. The fit is carried out by minimising the $\chi^2$ between data and theory using the \texttt{Ceres Solver} package~\cite{Agarwal_Ceres_Solver_2022}, and the fit uncertainties are estimated using the criterion of $\Delta\chi^2=1$ using the \texttt{HESSE} algorithm in the \texttt{MINUIT} package~\cite{James:1975dr}.
The $\chi^2$ definition follows that of Eq.~(32) in Ref.~\cite{H1:2015ubc}. 
For the starting scale $Q_0$ and virtuality cut $Q^2_{\min}$, two setups are considered: (1) $Q_0 = 1.6$~GeV, $Q^2_{\min} = 3.5$~GeV$^2$, and (2) $Q_0 = 3.2$~GeV, $Q^2_{\min} = 10$~GeV$^2$. The charm-threshold matching scale is set to the value of $Q_0$.

\section{Impact on PDFs}
\label{sec-pdf}

\begin{figure}
	\begin{minipage}{0.49\linewidth}
		{\includegraphics[width=1.0\linewidth]{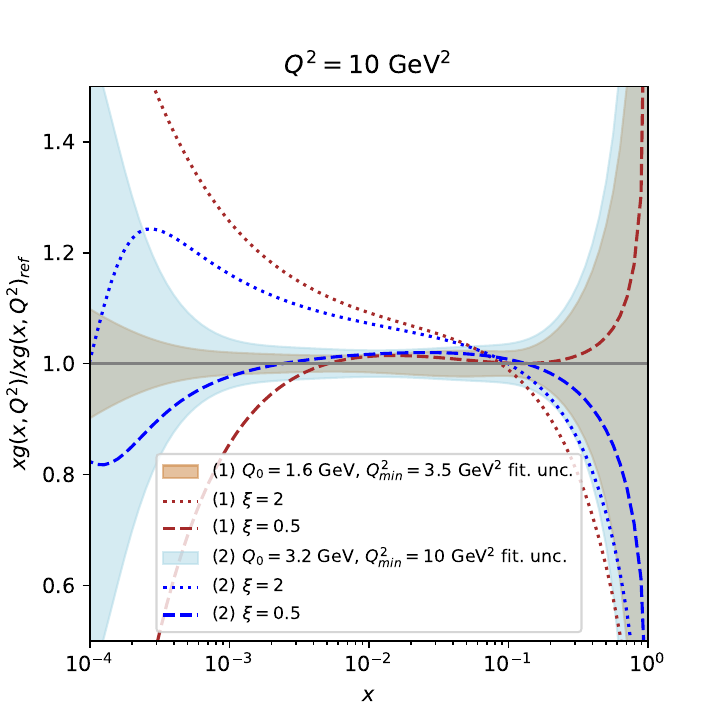}}
	\end{minipage}
	\begin{minipage}{0.49\linewidth}
		{\includegraphics[width=1.0\linewidth]{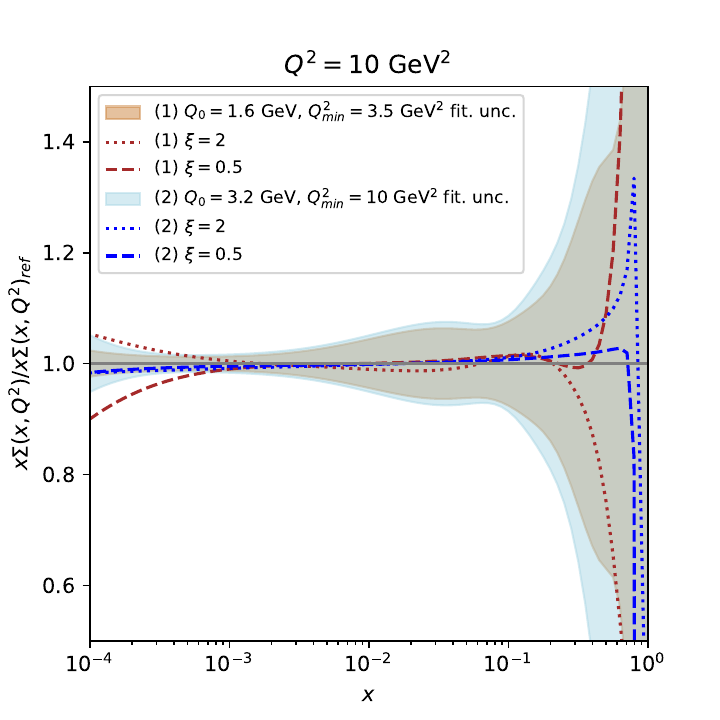}}
	\end{minipage}
	\caption[]{The fitted gluon and sea quark PDFs at $Q^2=10$ GeV$^2$ obtained in the fits with $Q_0=1.6$ GeV, $Q^2_{\min} = 3.5$ GeV$^2$, and $Q_0=3.2$ GeV, $Q^2_{\min} = 10$ GeV$^2$ with $\xi=0.5$, $\xi=1$ and $\xi=2$. The results are normalised to the PDFs obtained with $\xi=1$, for which the fit uncertainties are shown also.}
	\label{fig:q2_10}
\end{figure}

\begin{figure}
	\begin{minipage}{0.49\linewidth}
		{\includegraphics[width=1.0\linewidth]{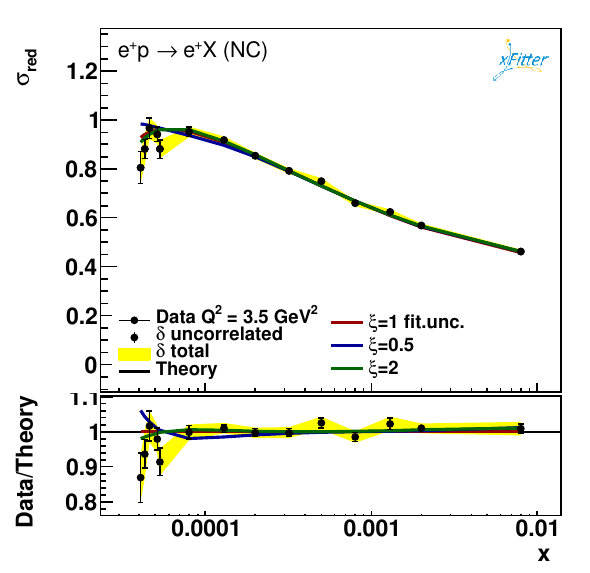}}
	\end{minipage}
    \begin{minipage}{0.49\linewidth}
		{\includegraphics[width=1.0\linewidth]{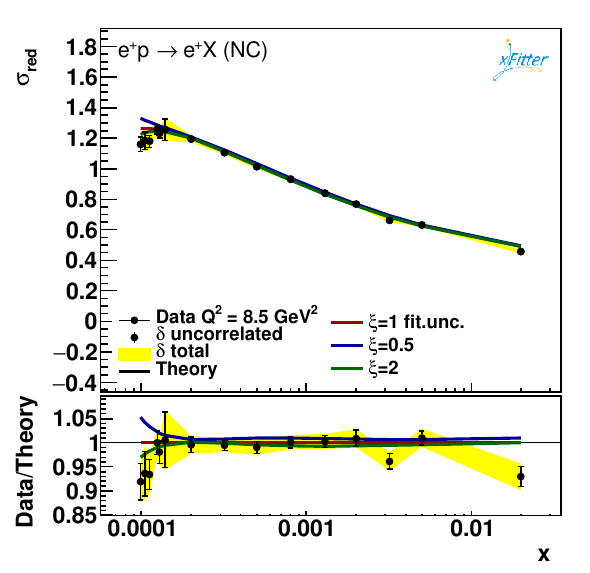}}
	\end{minipage}
    \caption[]{Data to theory comparison for a representative set of the HERA DIS NC reduced cross sections. In the bottom panels the ratios theory/data for the nominal variant of the fit and the scale variations are shown.}
	\label{fig:data}
\end{figure}

\begin{figure}
	\begin{minipage}{0.49\linewidth}
		{\includegraphics[width=1.0\linewidth]{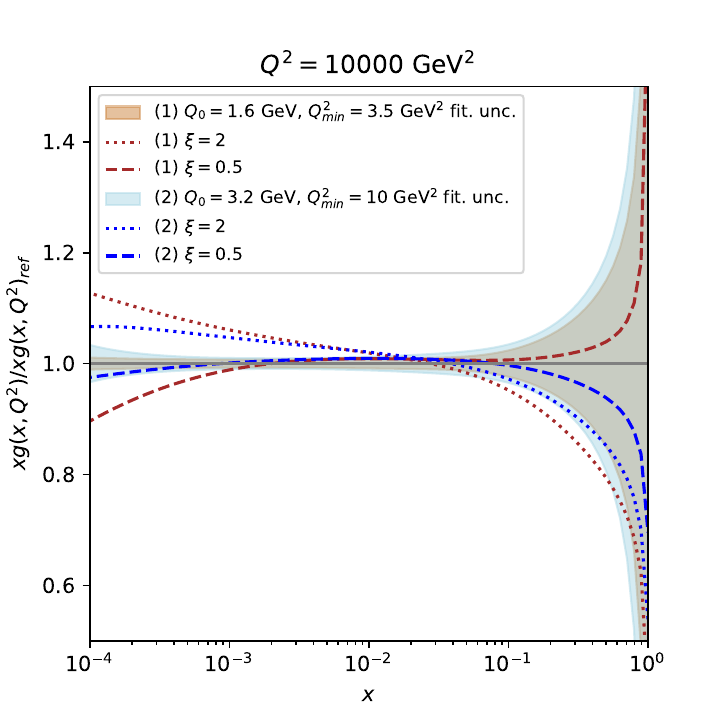}}
	\end{minipage}
	\begin{minipage}{0.49\linewidth}
		{\includegraphics[width=1.0\linewidth]{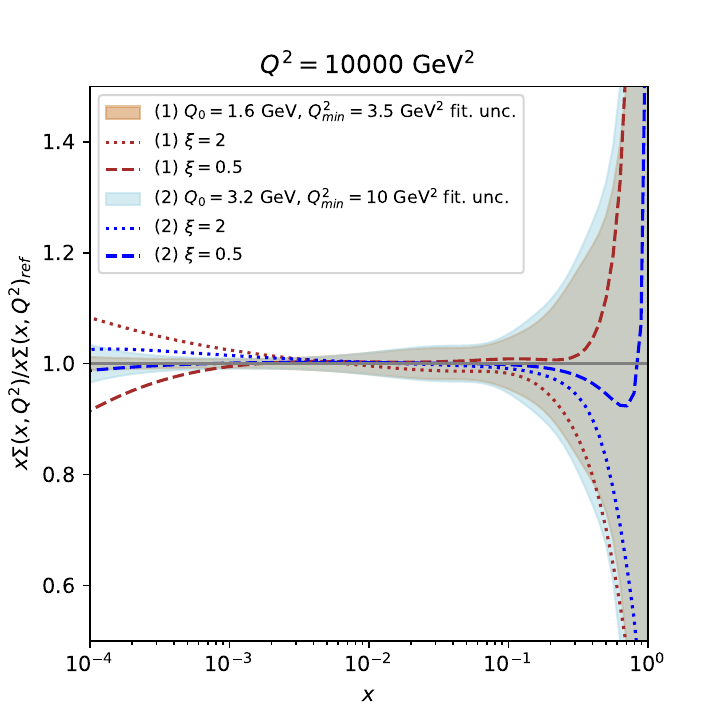}}
	\end{minipage}
	\caption[]{Same as in Fig.~\ref{fig:q2_10} for $Q^2=10000$ GeV$^2$.}
	\label{fig:q2_10000}
\end{figure}

The fitted PDFs at the energy scale $Q^2=10$ GeV$^2$ obtained with $Q_0=1.6$ GeV and $Q^2_{\min} = 3.5$~GeV$^2$ are shown in Fig.~\ref{fig:q2_10}.
With $\xi=1$, we obtain $\chi^2$ per degree of freedom $\chi^2/\text{dof} = 1312/1127$ (consistent with Ref.~\cite{Bonvini:2019wxf}).
A significant sensitivity to resummation-scale variations is observed in the fitted PDFs.
In particular, the gluon distribution and the sea quark distribution show variations significantly exceeding their fit uncertainties over a wide range of $x$.
The corresponding impact on the HERA DIS NC reduced cross sections for a representative set of data in the low-$Q^2$ and low-$x$ region is shown in Fig.~\ref{fig:data}. In this kinematic region higher-order corrections are expected to be important. In particular,  
logarithmically-enhanced contributions to splitting functions  
 are present at all orders for low $x$~\cite{Jaroszewicz:1982gr,Catani:1993ww,Catani:1993rn}. 
The observed impact on the PDFs persists after evolution to high scales $Q^2 \sim 10000$ GeV$^2$, as shown in Fig.~\ref{fig:q2_10000}.

The sensitivity to the resummation scale can be reduced by modifying the fit setup.
Since the largest deviations are observed in the low-$Q^2$ and low-$x$ region, the sensitivity to resummation-scale variations is expected to decrease if the low-$Q^2$ region is excluded from the fit.
Furthermore, resummation-scale variations produce larger shape variations of the PDFs at lower evolution scales, which may be more difficult to accommodate within a given PDF parametrisation. It can therefore be beneficial to choose a higher starting scale $Q_0$, although this does not in itself reduce the RGE uncertainty when the PDFs are consistently refitted.
The effects can be quantified using the logarithmically weighted average uncertainty of the gluon PDF for various fit configurations. Increasing $Q^2_{\min}$ removes data that are particularly sensitive to the gluon PDF at low $x$, thereby increasing the experimental uncertainty, while reducing the sensitivity to resummation-scale variations. The interplay between the experimental and theoretical uncertainties motivates the choice $Q_0=3.2~\mathrm{GeV}$ and $Q^2_{\min}=10~\mathrm{GeV}^2$ for the alternative fit setup.

The results of the fit using $Q_0 = 3.2~\mathrm{GeV}$ and $Q^2_{\min} = 10~\mathrm{GeV}^2$ are shown in Fig.~\ref{fig:q2_10}. This choice leads to a substantial reduction in the resummation-scale uncertainty, since the latter is driven largely by the low-scale region, where perturbative convergence is less stable.
Even though this reduction comes at the price of increased fit uncertainties of the PDFs in our case study, in global PDF fits high-energy collider data would compensate for reduced DIS statistics and enable a robust extraction of the PDFs~\cite{Alekhin:2025qdj}. In this variant of the fit, we obtain $\chi^2$ per degree of freedom $\chi^2/\text{dof} = 1124/998$ when using $\xi=1$.

\begin{figure}
	\begin{minipage}{0.49\linewidth}
		{\includegraphics[width=1.0\linewidth]{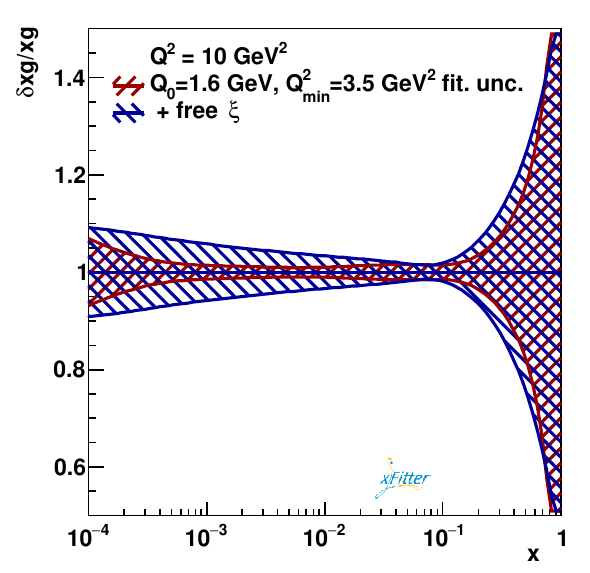}}
	\end{minipage}
	\begin{minipage}{0.49\linewidth}
		{\includegraphics[width=1.0\linewidth]{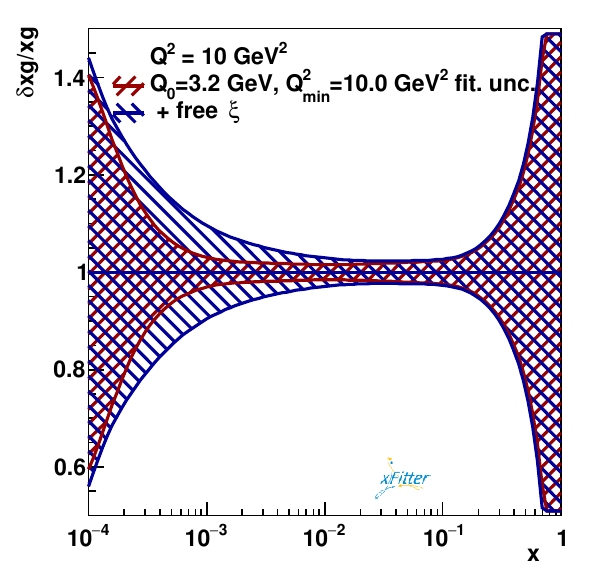}}
	\end{minipage}
	\caption[]{The gluon PDF fit uncertainties at $Q^2=10$ GeV$^2$ obtained in the fit with and without free $\xi$ parameter using $Q_0=1.6$ GeV, $Q^2_{\min} = 3.5$ GeV$^2$ (left), and $Q_0=3.2$ GeV, $Q^2_{\min} = 10$ GeV$^2$ (right).}
	\label{fig:q2_10_fitxi}
\end{figure}

An alternative way to account for the resummation-scale variation uncertainty is to treat the resummation scale $\xi$ parameter as a free parameter in the PDF fit, allowing it to be constrained by the data. This method is similar to the covariance-matrix approach for theoretical uncertainties used e.g.\ in Refs.~\cite{NNPDF:2019ubu,NNPDF:2024dpb}. Allowing $\xi$ to vary leads to a noticeable increase of the PDF uncertainties, most prominently for the gluon and sea-quark distributions. The resulting gluon PDF fit uncertainties are compared with the conventional fit uncertainties obtained by varying only the PDF parameters in Fig.~\ref{fig:q2_10_fitxi}. The fitted values of the resummation-scale parameter are $\xi = 2.1 \pm 0.3$ for the fit with $Q_0=1.6~\mathrm{GeV}$ and $Q^2_{\min}=3.5~\mathrm{GeV}^2$, and $\xi = 0.7 \pm 0.2$ for the fit with $Q_0=3.2~\mathrm{GeV}$ and $Q^2_{\min}=10~\mathrm{GeV}^2$. These values of $\xi$ indicate that the preferred scale can differ from the conventional choice $\xi=1$, depending on the fit setup. Furthermore, the ranges of $\xi$ constrained by the data appear to be much narrower than the conventional $0.5 < \xi < 2$ variation range. As a result, the increase of the PDF uncertainties due to the resummation-scale variations is generally smaller than in Fig.~\ref{fig:q2_10}. Nevertheless, the same qualitative effect is observed, i.e.\ the resummation-scale uncertainty is reduced when $Q_0 = 3.2~\mathrm{GeV}$ and $Q^2_{\min} = 10~\mathrm{GeV}^2$ are used in the fit.

\section{Impact on LHC observables}
\label{sec-lhc}

The resummation-scale uncertainty propagates to predictions for collider observables. As an example, we computed the top-quark pair production cross section at NNLO QCD at center-of-mass energies of $\sqrt{s}=5, 13$ TeV for the LHC, and $100$ TeV  for the FCC, using the \texttt{Hathor} library~\cite{Aliev:2010zk} interfaced in \texttt{xFitter}. 

\begin{figure}
	\centering
	\begin{minipage}{0.75\linewidth}
		\includegraphics[width=1.0\linewidth]{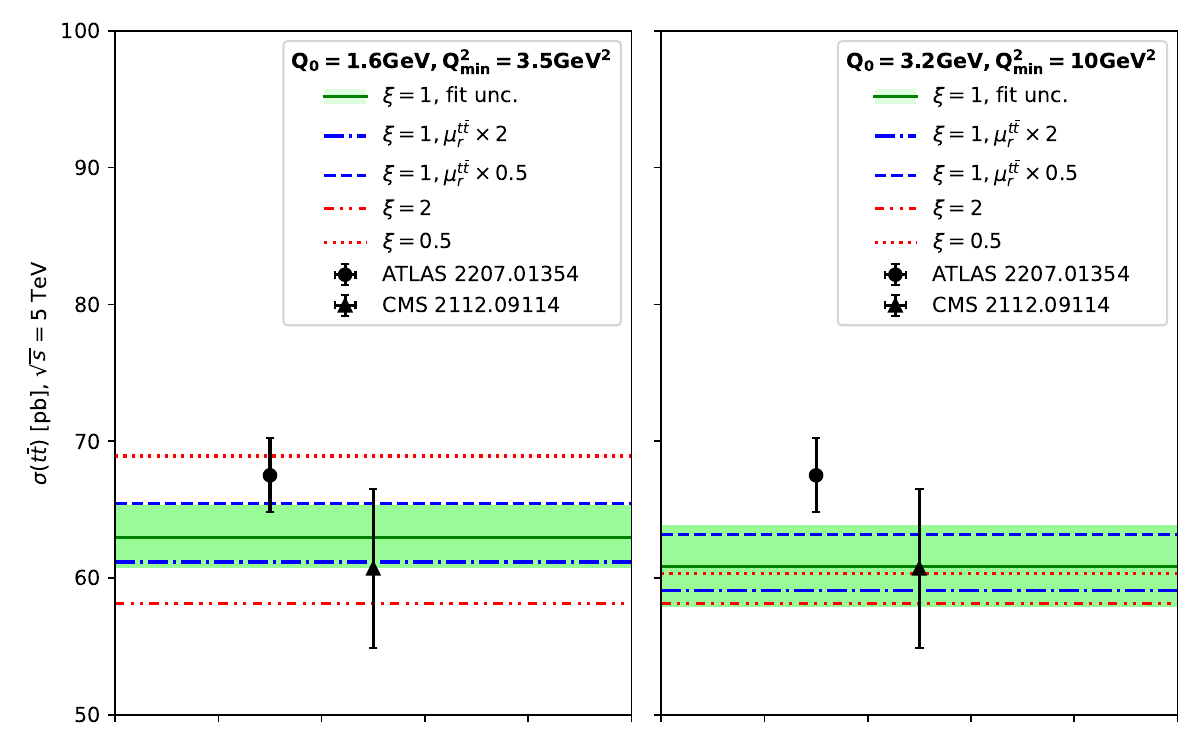}
	\end{minipage}
	\begin{minipage}{0.75\linewidth}
		\includegraphics[width=1.0\linewidth]{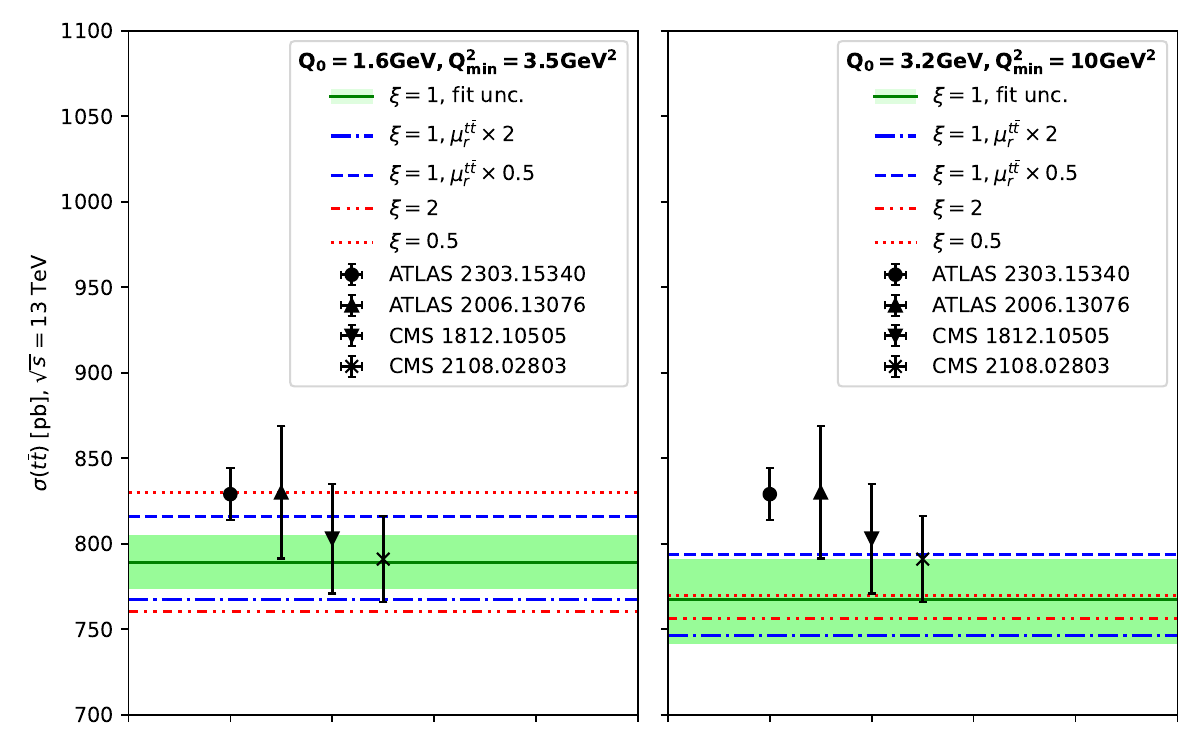}
	\end{minipage}
	\begin{minipage}{0.75\linewidth}
		\includegraphics[width=1.0\linewidth]{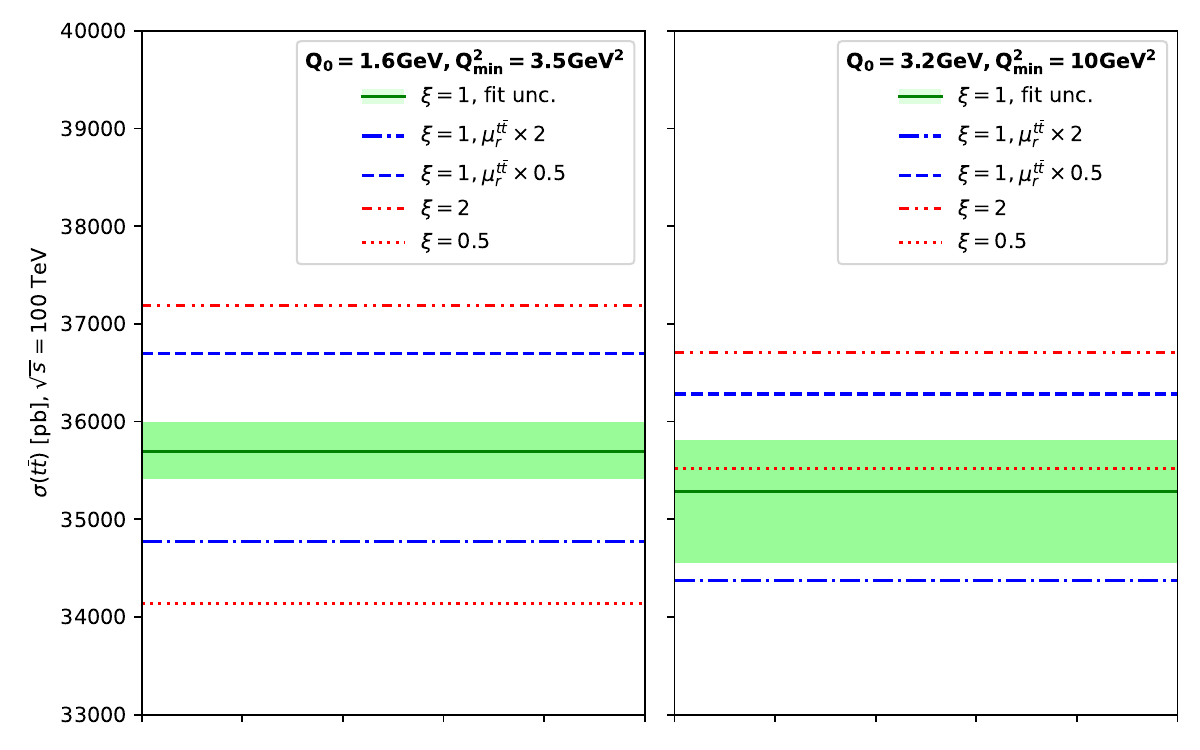}
	\end{minipage}
	\caption[]{$t\bar{t}$ cross sections at the LHC at $\sqrt{s} = 5$ TeV (upper) and $\sqrt{s} = 13$ TeV (middle), and at the FCC at $\sqrt{s} = 100$ TeV (lower) predicted at NNLO using the fitted PDFs. The cross sections are shown together with the PDF fit uncertainties and $\mu_r$ variation uncertainties, and resummation-scale variation ($\xi$) uncertainties. Data from ATLAS~\cite{ATLAS:2023gsl,ATLAS:2020aln} and CMS~\cite{CMS:2018fks,CMS:2021vhb} are also shown.}
	\label{fig:tt13}
\end{figure}

For this computation, we use the PDFs obtained from our fit to the HERA DIS data, together with their fit uncertainties and the PDF sets corresponding to resummation-scale variations by a factor of two up and down. The latter variations affect the top-quark pair production cross section at the level of $\mathcal{O}(5\%)$, which is comparable to the current experimental uncertainties and to the conventional renormalisation scale variation%
\footnote{The corresponding uncertainty arising from the variation of the factorisation scale is much smaller and is therefore not taken into account.}
, as shown in Fig.~\ref{fig:tt13}. This demonstrates that the uncertainty associated with the resummation procedure constitutes an additional source of theoretical uncertainty in precision phenomenological studies based on DIS PDF fits. Since top-quark pair production is predominantly sensitive to the gluon PDF at intermediate and large values of $x$, the observed effect illustrates how uncertainties originating from the treatment of the low-$Q^2$ DIS region propagate through the PDF determination and affect precision LHC predictions. However, with the improved PDF fit setup using $Q_0=3.2$ GeV and $Q^2_{\min}=10$ GeV$^2$, this uncertainty is reduced to approximately $1\%$, demonstrating that an appropriate choice of the fit settings can largely suppress the sensitivity to the resummation scale. At $\sqrt{s}=100$ TeV, the resummation-scale uncertainty is also reduced with the improved fit setup, although it remains at the few-percent level.

Furthermore, it is also worth noting that the reduction in the resummation-scale uncertainty is more pronounced at $\sqrt{s}=5$ TeV than at $\sqrt{s}=13$ and/or $\sqrt{s}=100$ TeV. At the lower center-of-mass energy, the top-quark pair production cross section is sensitive to the gluon distribution at larger values of $x$. This region is not directly constrained by the HERA data, but rather by the PDF extrapolation, which also depends on the momentum sum rule and therefore on the PDF shape at small $x$, namely the region most strongly affected by the resummation-scale variations.

\section{Conclusions}
\label{sec-conc}

For the first time, we have studied the impact of RGE theoretical uncertainties in PDF fits. This has been done  by employing the method of
resummation-scale variations and performing an
NNLO QCD fit to HERA DIS data using \texttt{xFitter}. 
This source of uncertainty provides a complementary estimate of missing higher-order effects in RGE evolution. 
These uncertainties propagate to LHC observables at a non-negligible level. This highlights the importance of including resummation-scale uncertainties in precision QCD analyses. 
Furthermore, we have shown that the impact of the resummation-scale uncertainty can be substantially reduced by excluding the low-$Q^2$ region from the fit.

\acknowledgments

The work of O. Z. has been supported by the \emph{MSCA4Ukraine Programme} of the European Commission through the Alexander von Humboldt foundation. The work of V.~B. has been supported by l’Agence Nationale de la Recherche (ANR), project ANR-24-CE31-7061-01.




\bibliographystyle{JHEP}
\bibliography{biblio} 

\newpage
\appendix
\section{Supplementary materials}

\begin{figure}[H]
	\centering
	\begin{minipage}{0.7\linewidth}
		\includegraphics[width=1.0\linewidth]{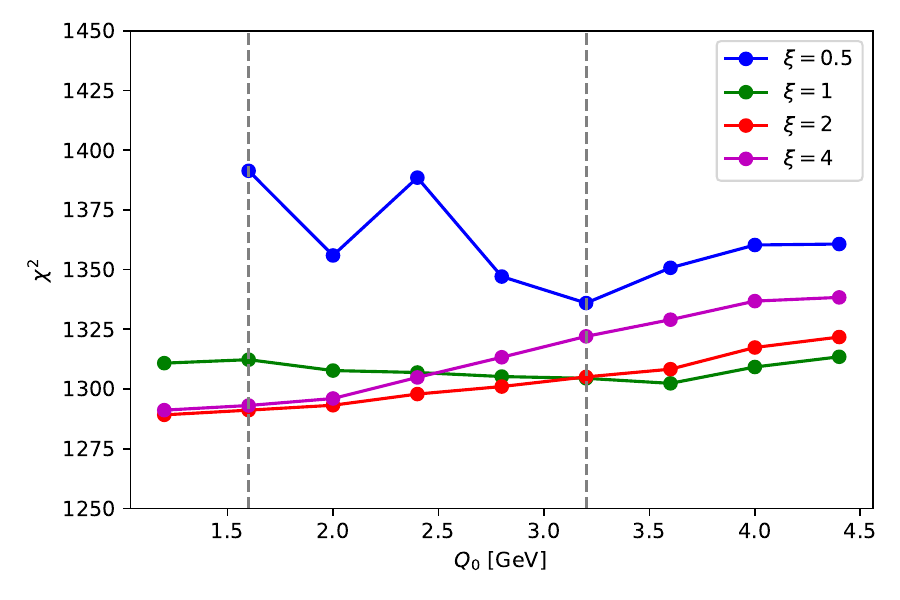}
	\end{minipage}
	\caption[]{A scan of $\chi^2$ as a function of $Q_0$ for $\xi=0.5$, $\xi=1$, $\xi=2$ and $\xi=4$.}
	\label{fig:q0scan}
\end{figure}

\begin{figure}[H]
	\centering
	\begin{minipage}{0.7\linewidth}
		\includegraphics[width=1.0\linewidth]{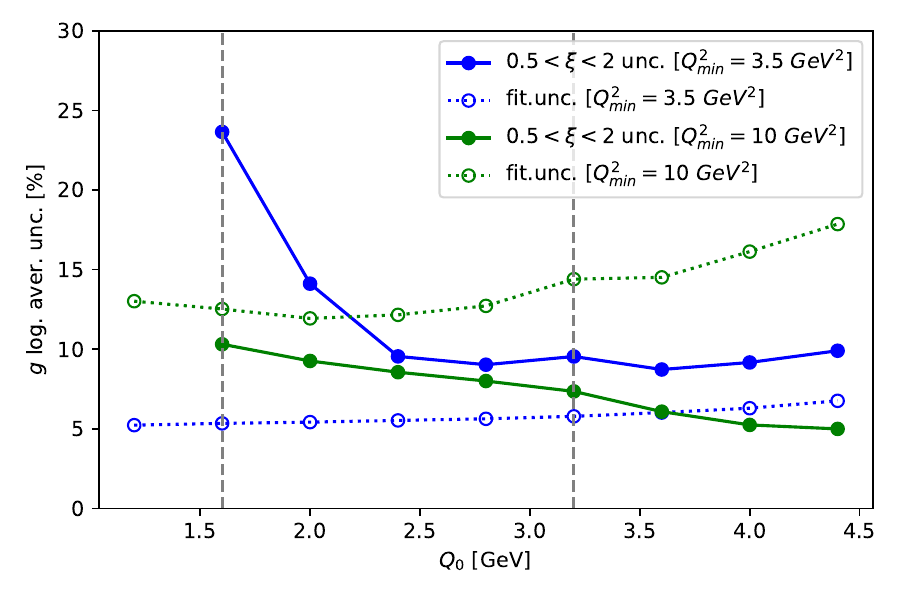}
	\end{minipage}
	\caption[]{The $g$ fit and resummation-scale variation uncertainty as a function of $Q_0$ at $Q^2=10$ GeV$^2$ for two variants of the fit with $Q^2_{\min} = 3.5$~GeV$^2$ and $Q^2_{\min} = 10$~GeV$^2$.}
	\label{fig:q0scan_pdfunc}
\end{figure}

A scan of the fit quality $\chi^2$ and the logarithmically weighted average uncertainty of the gluon PDF for various fit configurations as a function of the starting scale \(Q_0\) is presented in Figs.~\ref{fig:q0scan} and \ref{fig:q0scan_pdfunc}, respectively.

\end{document}